\documentclass[]{elsarticle} 
\biboptions{sort&compress}

\usepackage{amssymb}
\usepackage{amsmath}
\usepackage{subfig}
\usepackage{graphicx}
\usepackage{color}

\journal{Physica A}

\begin{document}
\title{Analysis of a decision model in the context of equilibrium pricing and order book pricing}

\author[ude]{D.C.~Wagner\corref{cor1}}

\author[ude]{T.A.~Schmitt\corref{cor2}}

\author[ude]{R.~Sch\"afer}

\author[ude]{T.~Guhr}

\author[ude]{D.E.~Wolf}

\cortext[cor1]{Corresponding Author: D.C.~Wagner; Email, daniel.wagner@uni-due.de; Phone, +49 203 379 4737}

\address[ude]{Faculty of Physics, University of Duisburg-Essen, Lotharstrasse 1, 47057 Duisburg, Germany}

\date{\today}

\begin{abstract}
An agent-based model for financial markets has to incorporate two aspects: decision making and price formation. We introduce a simple decision model and consider its implications in two different pricing schemes. First, we study its parameter dependence within a supply-demand balance setting. We find realistic behavior in a wide parameter range. Second, we embed our decision model in an order book setting. Here we observe interesting features which are not present in the equilibrium pricing scheme. In particular, we find a nontrivial behavior of the order book volumes which reminds of a trend switching phenomenon. Thus, the decision making model alone does not realistically represent the trading and the stylized facts. The order book mechanism is crucial.
\end{abstract}


\maketitle

\section{Introduction}
A common method in the economic literature to determine the price of an asset is the concept of equilibrium pricing~\cite{MasColell1995,Dixon2001}. The price is the result of the available supply and demand. In the context of stock markets supply and demand are often identified with the market expectation of the traders to sell or buy a stock. For example, such a pricing scheme is frequently employed within Ising-type models~\cite{Bornholdt2001,Kaizoji2002,Kim2008}, where the market expectation of each trader is symbolized by a spin on a lattice. There the decision to sell or buy a stock is mapped to spin down and up. In each time step an update of the market expectations is performed by taking the market expectations of the nearest neighbors into account~\cite{Bornholdt2001}. A possible criterion to derive the equilibrium price is to take the average of all market expectations.

In contrast, the standard pricing mechanism at stock exchanges is the double auction order book~\cite{Friedman1993,Johnson2010}. The order book lists all current propositions to sell or buy at a given price. This information is available to all traders. Here, the price is the result of a new incoming order matching a limit order already in the book.

The empirical return time series show a collection of remarkable properties, which are called stylized facts, see Ref.~\cite{Chakraborti2011} for a review. The empirical distribution of returns has heavy tails, \textit{i.e.}, the tails are more pronounced compared to a normal distribution~\cite{Mandelbrot1963,Plerou1999,Mantegna1999,Plerou2004}. In the literature many approaches are discussed to explain this feature~\cite{Clark1973,Arthur1996,Mandelbrot1997,Sornette1998,Lux1999,Challet2000,Farmer2004,Gabaix2003,Farmer2004a}. One point of view  claims that large trading volumes are responsible for the heavy tails~\cite{Clark1973,Arthur1996,Mandelbrot1997,Sornette1998,Lux1999,Challet2000}. Another approach holds gaps in the order book structure responsible. If gaps are present between the price levels occupied by limit orders, even small volumes can cause large price shifts~\cite{Farmer2004}. According to this reasoning the order book plays an important role in the emergence of heavy tails. 

Agent-based modeling makes it possible to microscopically understand trading mechanisms~\cite{Cohen1983,Kim1989,Frankel1988,Chiarella1992,Beltratti1993,Levy1994,Lux1997,Mike2008,Gu2009} and to study trading strategies~\cite{LeBaron2002,Ehrentreich2007}. Here, we extend the agent-based stock market introduced by Schmitt \textit{et al.}~\cite{Schmitt2012} which implements a double auction order book. The model is capable of reproducing the gap structure described in Ref.~\cite{Farmer2004} and yields heavy-tailed return distributions. The traders employed in the model act randomly, independently of each other and follow no strategy.

We extend this model by designing a trader that acts with respect to a decision model. The traders are indirectly coupled to each other because their decisions influence the price which, in turn, affects their decision to buy or to sell an asset, see Sec. 2. Since the model has two parameters we initially analyze the parameters' influence on the price with an equilibrium pricing. Then we analyze the persistence of this behavior if we put the traders' decision model into an order book setting. As results of our simulations we discuss the return distribution and the order book structure in Sec. 3. We conclude our results in Sec. 4.

\section{Model}
We present a decision model in an equilibrium pricing and discuss how to implement it in an order book driven agent-based model. In reality, traders employ all kinds of strategies. Unfortunately, it is not possible to capture the decision making process of any trader. Furthermore, his strategy could be erratic or could at least contain erratic elements, so that it would be impossible to exactly determine how he would react in a particular case. Also, the more complex a system is, the less it is possible to trace back features in the observables, \textit{e.g.} the price, to individual decisions. Therefore we restrict ourselves to a simple decision model where this is still possible.

In an agent-based model for a financial market the decisions of every trader (agent) and their influence on the price are simulated. Thus, agent-based models have to take into account two different aspects: decision making and price formation. A decision model describes how a trader reacts on events, \textit{e.g.} when the price changes, and price formation is a mechanism describing how decisions manifest themselves in the price. Here we analyze one decision model in combination with two different pricing models:

Our decision model demands that every trader has his own price estimation for a traded asset. Of course this influences his market expectation and, thus, his decision either to buy or sell. Depending on the current asset price his individual price estimation develops in time. He changes his decision to buy or to sell, respectively, depending on the relative deviation between the individual and the asset price; a distribution function determines the probability that he changes his market expectation.

The first pricing model calculates the relative price change between two time steps as the mean value of the market expectations of all traders. This corresponds to an equilibrium pricing, balancing supply and demand. In the second price formation scheme the price is determined in the framework of an order book setting. Here we embed the decision model into an agent-based model with a double auction order book pricing. We will now look deeper into these aspects.

\subsection{Decision making}
By their very nature decision making models are capable of describing a large variety of scenarios, wherever a choice between alternatives has to be made. Contrary to approaches in the literature where the decisions are based on mutual decisions between the agents~\cite{Torok2013}, the decisions in the present setting are only made with the reference to a general trend, which can be understood as a mean field.

Let us consider $i=1,\dots ,N$ agents who order to buy or offer to sell a certain asset. Only a single asset is considered. Its supply, quality, usefulness etc. are not taken into account. Whether an agent orders to buy or offers to sell depends on his market expectation $m_i(t)$ about the current price. If he thinks the asset is overpriced $m_i(t)=-1$ he offers to sell, if he thinks it is more worth than the current price $m_i(t)=1$, he buys. To determine what an agent regards as the appropriate price constitutes the crucial part of the model. For simplicity it is assumed that every agent follows the price evolution from the time instant on when he last changed his market expectation. We call this time ``individual reset time'' $\tau_{i,j}$, where $j$ just enumerates those instants of time. What an agent thought before that instant is forgotten. Between individual reset times the individual market expectation $m_i(t)$ is constant.

At the reset time an agent $i$ changes his market expectation because at this time instant agent he felt the asset was more mispriced than he would tolerate. A simple ansatz for what he regards as the appropriate price at this time instant is
\begin{align}
s_i(\tau_{i,j}) = {S}(\tau_{i,j}) (1+m_i(\tau_{i,j})) \ .
\label{eq:reset}
\end{align}
The function $S(t)$ describes the time development of the asset price. Here, $S(\tau_{i,j})$ is the current price at the individual reset time $\tau_{i,j}$. We distinguish the quantity $s_i$ which is used to assess the benefit of the trader's market expectation with reference to the present price from the quantity $m_i$ which is a binary measure for the trader's market expectation. Eq.~\eqref{eq:reset} is a rather bold ansatz. Its advantage is the absence of free parameters. It means that the asset becomes worthless to the agent $s_i(\tau_{i,j})=0$, if he decides that he wants to sell $m_i(\tau_{i,j})=-1$. If he decides to buy, however, he thinks the appropriate price may be as much as twice as high as the current price $s_i(\tau_{i,j}) = 2 {S}(\tau_{i,j})$.

As long as agent $i$ does not change his market expectation again, the price he thinks appropriate evolves during a discrete time step $\Delta t$ according to
\begin{align}
s_i(t+\Delta t) = (1-\alpha)s_i(t) 
+ \alpha {S}(t+\Delta t) \ .
\end{align}
The parameter $\alpha \in [0,1]$ quantifies how readily the agent adjusts his estimation what the appropriate price should be compared to the actual price, taking the trend into account which explains why we use $S(t+\Delta t)$ instead of $S(t)$ on the right-hand side. Trivial cases are $\alpha=0$ (no adjustment) and $\alpha=1$ (current price always regarded as appropriate).

The last ingredient of the model concerns the condition that leads agent $i$ to change his market expectation. It depends on the relative deviation $h_i(t)$ of the current price from the one he regards as appropriate,
\begin{align} 
h_i(t) = \frac{s_i(t)-{S}(t)}{{S}(t)} \ .
\end{align}
If $m_i(t) h_i(t)>0$, there is no incentive to change. Either the actual price is lower than what is regarded as appropriate and the agent orders to buy, or the actual price is higher than what is regarded as appropriate and the agent wants to sell. However, if $m_i(t) h_i(t)<0$ the agent must consider to change his market expectation in order to avoid losses. On the other hand, a small mispricing might be a transient fluctuation which the agent may tolerate for a while without changing his market expectation. This means that there is a certain inertia in the variables $m_i(t)$. 

\begin{figure}[htbp]
  \begin{center}
    	\includegraphics[width=0.75\textwidth]{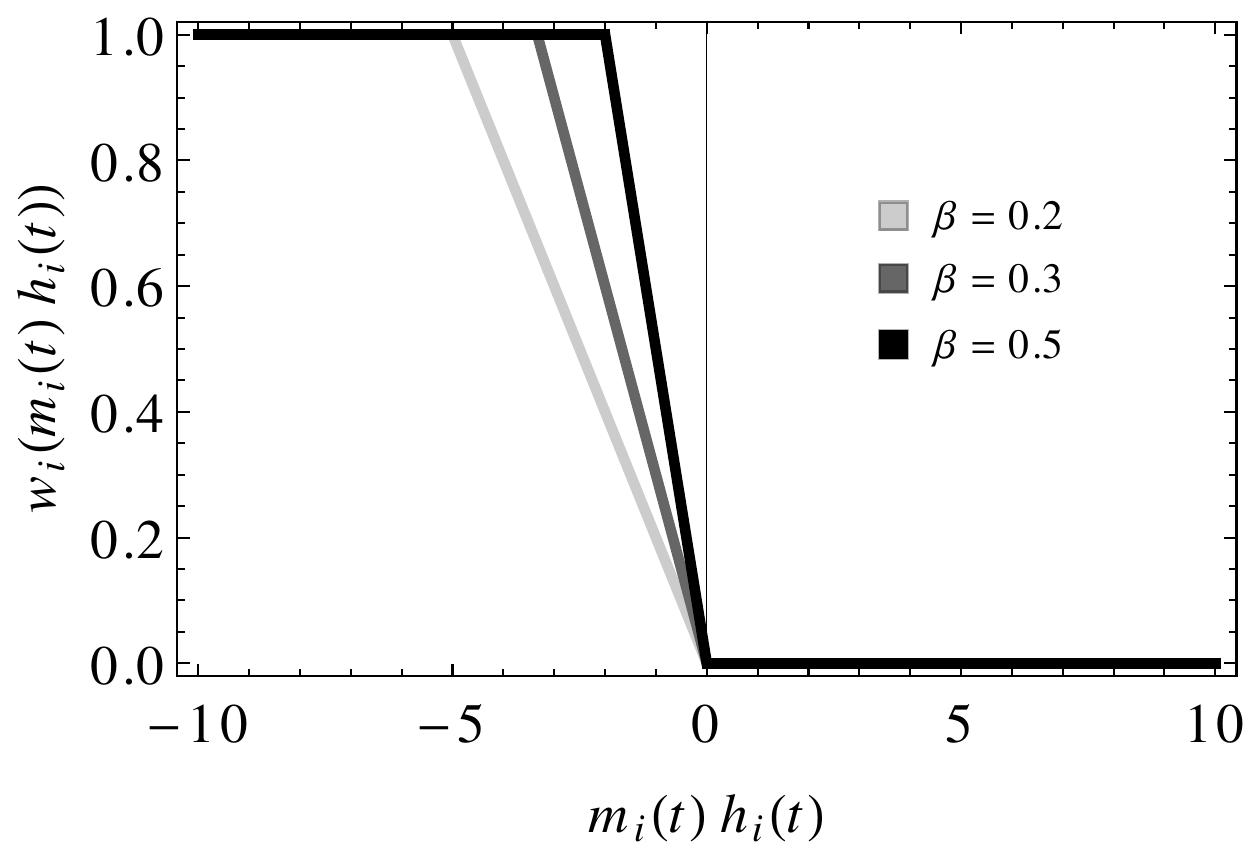}
  \end{center}
 \caption{Comparison of the changing probability Eq.~\eqref{eq:rate} for several values of $\beta$}
 \label{fig:dmm:rate}
\end{figure}

In this model, changing ones market expectation happens with a certain probability $w_i(m_i(t)h_i(t))$ which approaches zero for $m_i(t)h_i(t) \rightarrow \infty$ and one for $m_i(t)h_i(t)\rightarrow -\infty$. A possible choice is
\begin{align}
 w_i(m_i(t)h_i(t)) = \min(1,-\beta m_i(t)h_i(t))
\label{eq:rate}
\end{align}
if $m_i(t)h_i(t)<0$ and zero otherwise, see Fig.~\ref{fig:dmm:rate}.

Here, $1/\beta$ has the meaning of tolerance of an agent, ranging from zero to infinity. The changing probability is asymmetric, that is, depending on $\beta$, the trader only changes his market expectation if $m_i(t) h_i(t)$ is less than zero. It is worth mentioning that for $\beta \rightarrow 0$ the trader never changes his market expectation.

Obviously, other market expectation switching rules can be constructed in the same spirit. We discovered that choosing a point symmetric function (around $(0,0.5)$, like a Fermi-Dirac distribution) always leads to an exponential decrease of the price for increasing $t$. Furthermore we found that for $w_i(0)=0.5$ any local imbalance, that is, there are temporarily more traders who want to buy than who want to sell or \textit{vice versa}, vanishes from one time step to the next.

\subsection{Equilibrium pricing}
For a simple price formation the relative price change is only determined by the mean value of the variables $m_i(t)$:
\begin{align}
\frac{{S}(t+\Delta t)-{S}(t)}{{S}(t)} = \frac{1}{N} \sum_{i=1}^N m_i(t).
\label{eq:return}
\end{align}
Furthermore $N$ should be odd, so that this quantity is never exactly zero.

\subsection{Order book pricing}
\begin{figure}[htbp]
  \begin{center}
    	\includegraphics[width=0.75\textwidth]{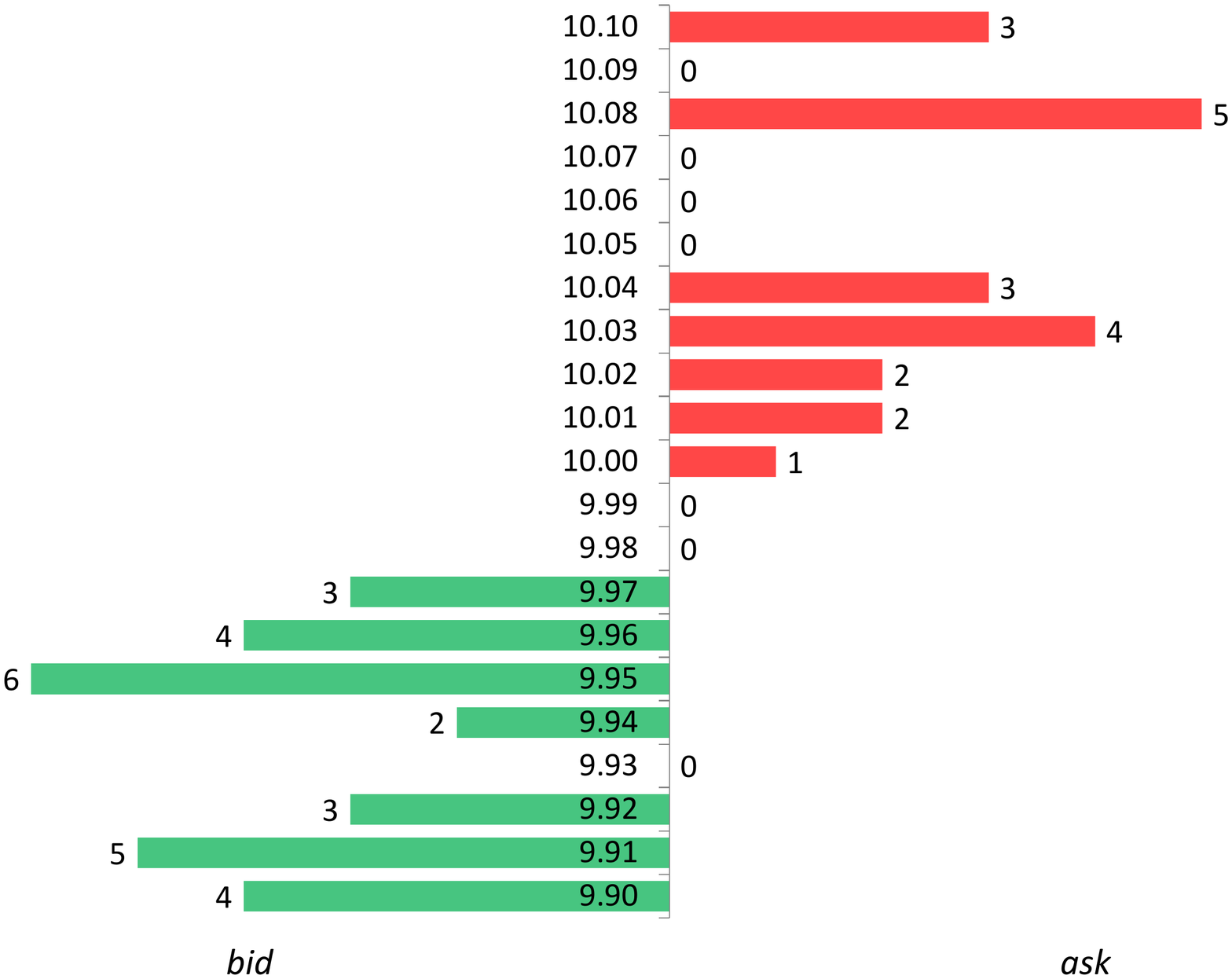}
  \end{center}
 \caption{Exemplary order book with bids/asks and their quantities}
 \label{fig:pm:order}
\end{figure}
At a stock exchange the clearing office manages an order book which is publicly available. The order book is filled by limit orders so that it contains the information about prices and volumes of all bids and asks. If the best (\textit{i.e.} lowest) ask is at least equal to the best (\textit{i.e.} highest) bid, a trade takes place. The traded price at this moment constitutes the current stock price. A trader decides if he wants to buy or sell a specific amount of stocks and he decides on the price he would agree with. In addition, the trader determines how long this limit order should exist in the order book at most. This is the so-called lifetime of an order. If he does not care about the price he is also able to submit a market order. Such a market order is executed immediately. As market orders remove limit orders, they enlarge the bid-ask spread in the order book and reduce liquidity.

In the exemplary order book in Fig.~\ref{fig:pm:order} there are gaps at the prices $9.93$, $10.05$ to $10.07$ and $10.09$. They can appear if limit orders vanish due to lifetime expiration or if no limit orders have been placed at that price level. If market orders remove enough volume, including multiple price levels with gaps, larger price shifts (\textit{i.e.} larger returns) occur.

\section{Results}
It is our goal to study how different pricing mechanisms, equilibrium or order book pricing, respectively, influence the observables considering our decision model.

\subsection{Equilibrium pricing}
We study our decision model using the equilibrium pricing in Eq.~\eqref{eq:return} where $50$ traders start with $m_i(0)=-1$ and $51$ with $m_i(0)=1$. The simulated price time series behave like a geometric Brownian motion for $\alpha=0.5$ and $\beta=0.5$. 

\begin{figure}[htbp]
  \begin{center}
    	\includegraphics[width=0.75\textwidth]{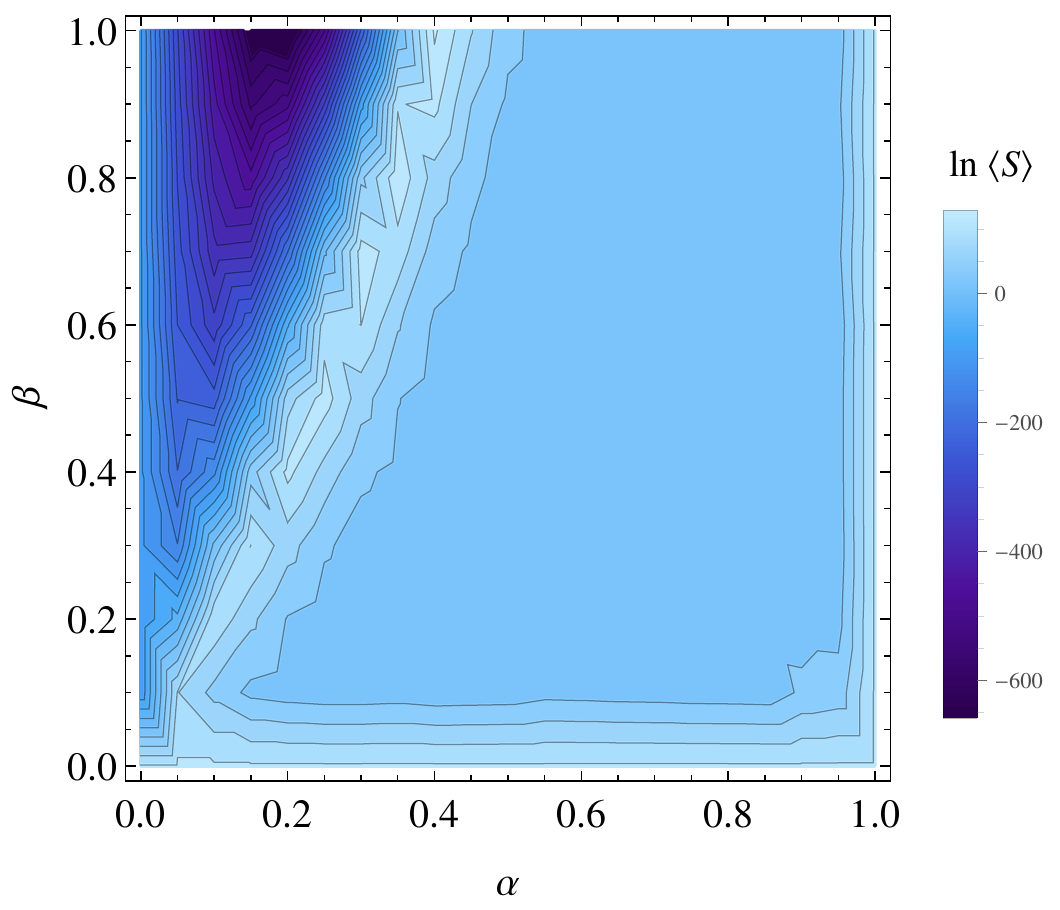}
  \end{center}
 \caption{Natural logarithm of the mean price $\langle S \rangle$ depending on $\alpha$ and $\beta$ as a mean value of $100$ seeds after $10^6$ time steps}
 \label{fig:results:B3D}
\end{figure}

Analyzing the dependencies on $\alpha$ and $\beta$, we find three regimes, see Fig.~\ref{fig:results:B3D}:
\begin{itemize}
 \item exponential increase of the price (strong positive drift),
 \item fluctuations of the price without a drift and
 \item exponential decrease of the price (strong negative drift).
\end{itemize}
In principle, $\alpha$ and $\beta$ determine the drift of the price time series. It turns out that there is a large parameter space ($\ln{\langle S \rangle}\approx 0$) in which the price neither diverges nor decays. This is important to generate sufficiently long time series for better statistics. We find that the returns are normal distributed. Given the equilibrium pricing the price time series of our decision model follows a geometric Brownian motion.

\subsection{Order book pricing}
In combination with the equilibrium pricing the price time series does not show any interesting features: Our simple decision model only yields a geometric Brownian motion. We are now going to investigate whether this is still the case in an order book setting.

We extend the agent-based model program by Schmitt \textit{et al.}~\cite{Schmitt2012} which implements a double auction order book. The hitherto used equilibrium pricing in Eq.~\eqref{eq:return} is no longer relevant, but for decision making the relation in Eq.~\eqref{eq:rate} is still used. Our model traders ($200$ of them are used in our simulations) only submit market orders. Their market expectations $m_i(t)$ determine if they submit a sell or a buy market order. In our simulations $400$ \texttt{RandomTrader}s act as liquidity providers; they fill the order book with limit orders whose values are randomly drawn from a normal distribution with a mean value of the current best bid or best ask, respectively. Their order lifetime influences the kurtosis of the return distribution drastically and is chosen in such a way that the return distribution is normal without any of our model traders, see~\cite{Schmitt2012}.

Our program adapts real trading behavior in so far as not every trader acts at every time step (here: one second). During $10^6$ seconds the traders each submit 1500 orders on average. In the decision model the individual price expectation $s_i$ updates every second. As time steps can be skipped in the course of the simulation, every time an agent trades we have to calculate the individual price expectation for the corresponding delay afterwards. Therefore we will see that $\alpha_\mathrm{abm} \sim \alpha / 1000$, that is, his price expectation in our agent-based model develops much more slowly than in our study with equilibrium pricing.

The criterion for changing the market expectation is often fulfilled within the time interval between two trades of an agent because of the previously mentioned missed time steps. Therefore in the agent-based model $\beta$ (inverse tolerance) is of minor importance. The simulations are almost independent of this parameter if it is within $[0.1, 1]$. For the results presented here, we chose $\beta=0.5$.

\begin{figure}[htbp]
  \begin{center}
    	\includegraphics[width=0.75\textwidth]{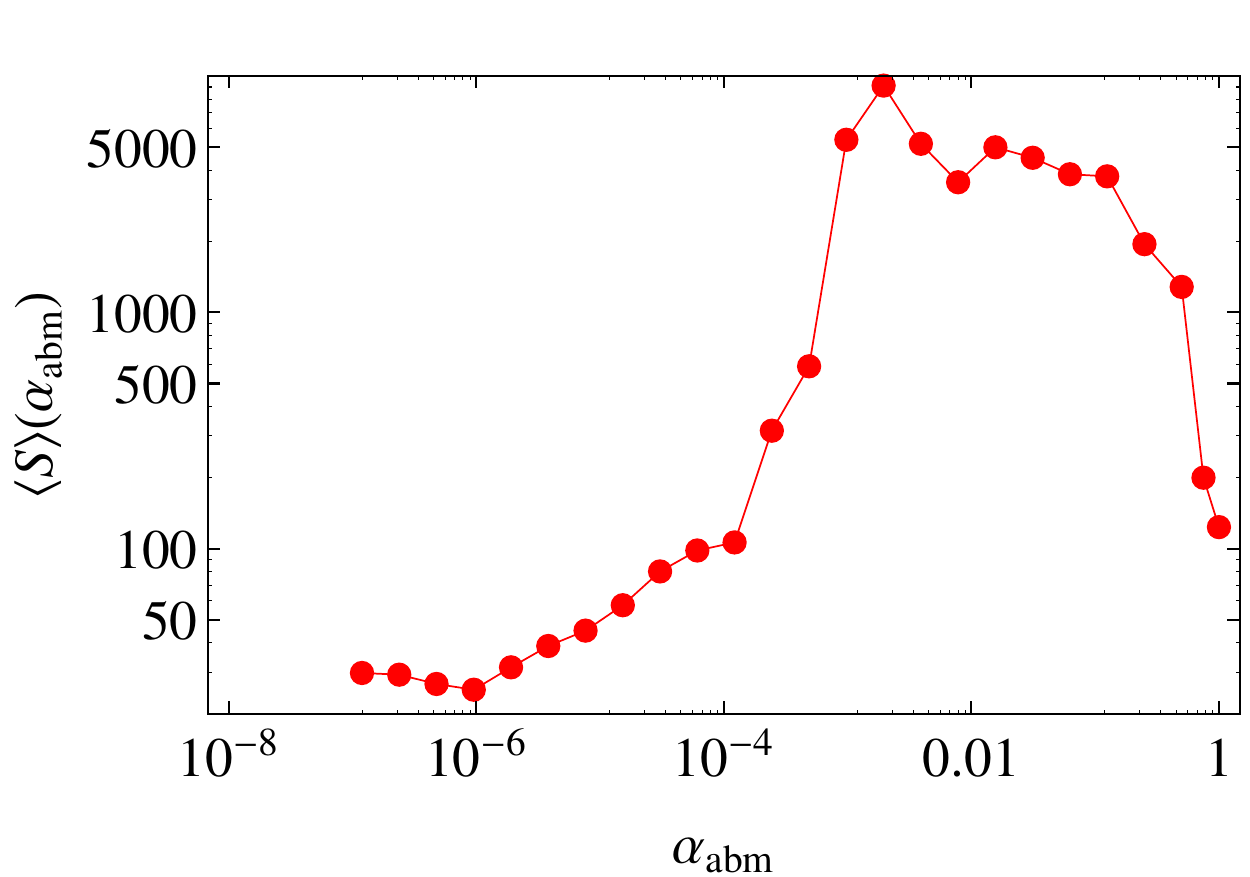}
  \end{center}
 \caption{Mean price $\langle S \rangle$ depending on $\alpha_\mathrm{abm}$ as a mean value of 100 seeds after $10^6$ time steps ($S(0)=100$ as starting value for every simulation)}
 \label{fig:results:abmalpha}
\end{figure}

\begin{figure}[htbp]
  \begin{center}
    	\includegraphics[width=0.75\textwidth]{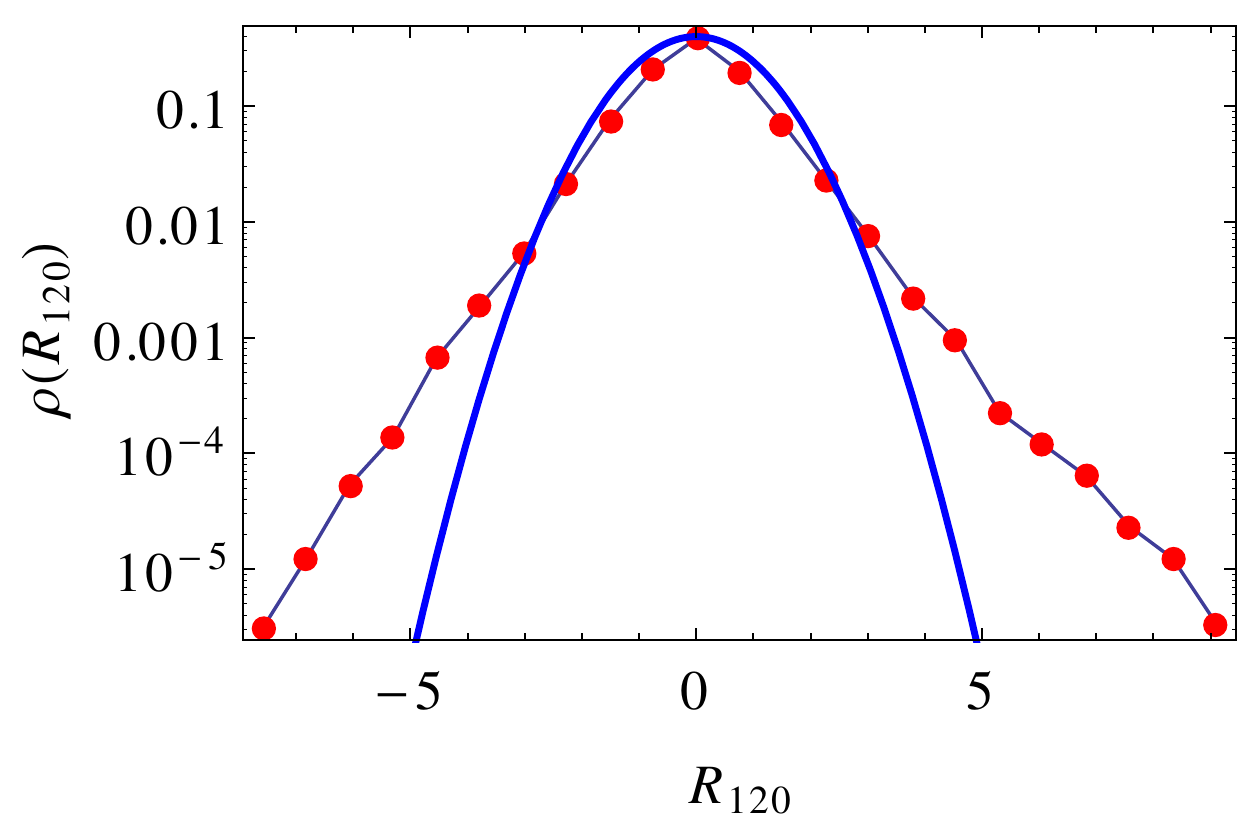}
  \end{center}
 \caption{Return distribution for $\alpha_\mathrm{abm}=10^{-4}$ with return interval $120\,\mathrm{s}$ and a normal distribution for comparison}
 \label{fig:results:abm4ret}
\end{figure}

Figure~\ref{fig:results:abmalpha} shows that there are also regimes where the price decreases, increases and fluctuates without a global drift.
Since the initial price is $S(0)=100$, we analyze $\alpha_\mathrm{abm}=10^{-4}$: The return distribution which we calculated for a return interval of $120\,\mathrm{s}$ is heavy-tailed with a kurtosis of about six, see Fig.~\ref{fig:results:abm4ret}. As already mentioned above, this shape can be modified by varying the lifetime of the limit orders.

\begin{figure}[htbp]
  \begin{center}
    	\includegraphics[width=0.75\textwidth]{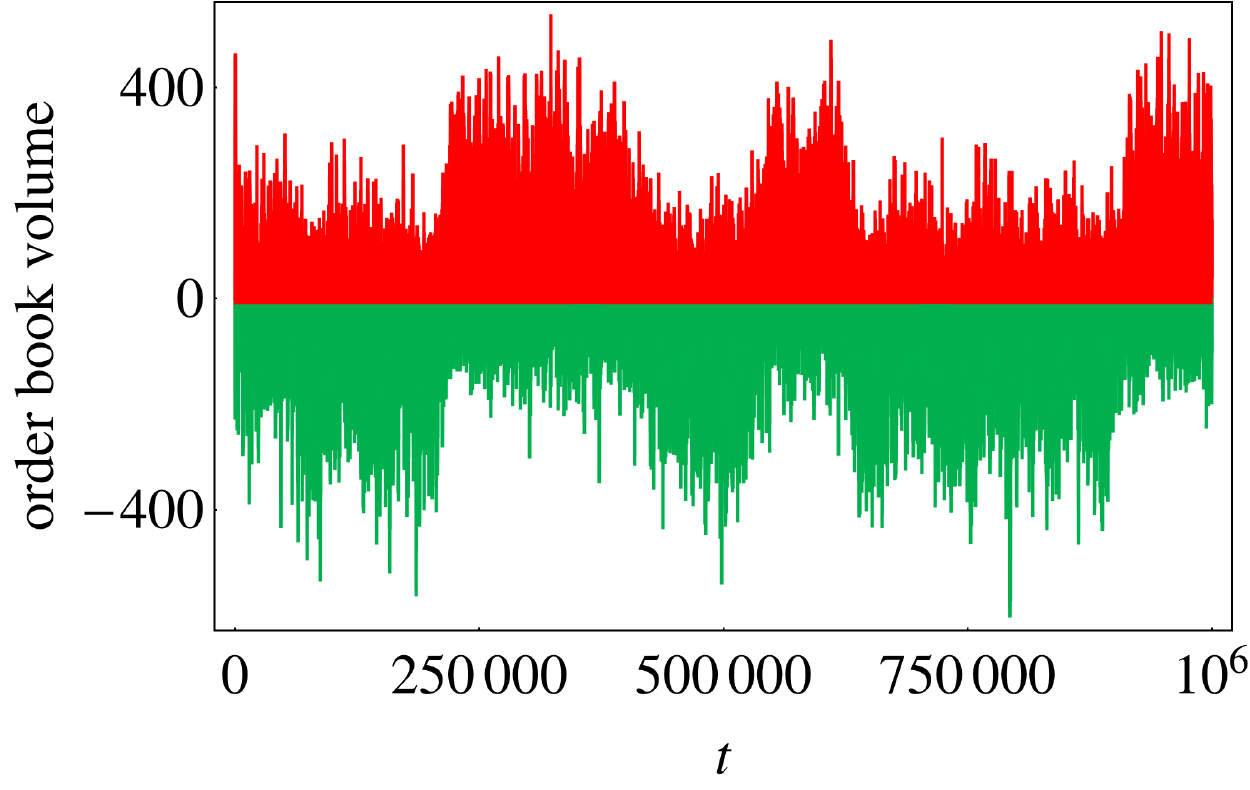}
  \end{center}
 \caption{Typical order book volume with sell (above) and buy orders (below)}
 \label{fig:results:abmobook}
\end{figure}

Another feature of this decision model is the herding behavior which can be observed on the order book volume, see Fig.~\ref{fig:results:abmobook}. There are time intervals in which much more sell orders are stored in the order book and time intervals in which much more buy orders are stored therein. This behavior is a feature of our decision model and cannot be reproduced by solely using $\mathtt{RandomTrader}$s.

\section{Conclusion}
For the equilibrium pricing we see a large space of parameters for which the model produces sustainable price time series, \textit{i.e.}, the price does not diverge or collapse. Heavy tails are not observed. The time series is reminiscent of a geometric Brownian motion where $\alpha$ and $\beta$ determine its global trend.

Embedding the decision model into an agent-based model with order book pricing causes the scale of the parameters to change. However, the parameter $\alpha$ of the decision model has comparable ramifications for the price time series. More importantly, we observe heavy tails for the return distribution. Thus, we conclude that an order book is necessary for the emergence of heavy tails in the context of our decision model. Combining a decision model with an order book pricing allows us to study the effects of the decision model within a realistic environment. We emphasize that in contrast to Schmitt \textit{et al.}~\cite{Schmitt2012} the lifetime of the \texttt{RandomTrader}s' limit orders is chosen so that near the best prices less gaps exist. At deeper price levels gaps do exist, but typically they play no role because the order volumes of the \texttt{RandomTrader}s do not dig deep enough into the order book.
The herding behavior of our model traders leads to a unidirectional stress to either the available ask or bid volume. By acting in the same direction the model traders are able to remove the densely occupied price levels around the best ask or bid and reach the gaps deeper in the order book. This is a crucial factor in generating heavy tails.

In general, implementing a decision model in the setting of an order book pricing leads to new insights and gives the possibility to study it in a realistic environment. The order book makes it possible to gain the understanding of a microscopic level and to access new features. Put differently, decision making models that are not embedded into an order book environment can yield unrealistic or even misleading results.

\section*{Acknowledgments}
We thank S\'ilvio R. Dahmen and Ana L.C. Bazzan for preliminary studies on the model.

\bibliographystyle{elsarticle-num-names}

\begin{thebibliography}{34}
\providecommand{\natexlab}[1]{#1}
\providecommand{\url}[1]{\texttt{#1}}
\providecommand{\urlprefix}{URL }
\expandafter\ifx\csname urlstyle\endcsname\relax
  \providecommand{\doi}[1]{doi:\discretionary{}{}{}#1}\else
  \providecommand{\doi}[1]{doi:\discretionary{}{}{}\begingroup
  \urlstyle{rm}\url{#1}\endgroup}\fi
\providecommand{\bibinfo}[2]{#2}

\bibitem[{Mas-Colell et~al.(1995)Mas-Colell, Whinston, and
  Green}]{MasColell1995}
\bibinfo{author}{A.~Mas-Colell}, \bibinfo{author}{M.~D. Whinston},
  \bibinfo{author}{J.~R. Green}, \bibinfo{title}{Microeconomic Theory},
  \bibinfo{publisher}{Oxford University Press}, \bibinfo{year}{1995}.

\bibitem[{Dixon(2001)}]{Dixon2001}
\bibinfo{author}{H.~D. Dixon}, \bibinfo{title}{Surfing Economics},
  \bibinfo{publisher}{Palgrave Macmillan}, \bibinfo{year}{2001}.

\bibitem[{Bornholdt(2001)}]{Bornholdt2001}
\bibinfo{author}{S.~Bornholdt}, \bibinfo{title}{{Expectation Bubbles in a Spin
  Model of Markets: Intermittency from frustration across scales}},
  \bibinfo{journal}{International Journal Of Modern Physics C}
  \bibinfo{volume}{12}~(\bibinfo{number}{5}) (\bibinfo{year}{2001})
  \bibinfo{pages}{667--674}.

\bibitem[{Kaizoji et~al.(2002)Kaizoji, Bornholdt, and Fujiwara}]{Kaizoji2002}
\bibinfo{author}{T.~Kaizoji}, \bibinfo{author}{S.~Bornholdt},
  \bibinfo{author}{Y.~Fujiwara}, \bibinfo{title}{{Dynamics of price and trading
  volume in a spin model of stock markets with heterogeneous agents}},
  \bibinfo{journal}{Physica A} \bibinfo{volume}{316} (\bibinfo{year}{2002})
  \bibinfo{pages}{441--452}.

\bibitem[{Kim et~al.(2008)Kim, Kim, and Yook}]{Kim2008}
\bibinfo{author}{Y.~Kim}, \bibinfo{author}{H.-J. Kim}, \bibinfo{author}{S.-H.
  Yook}, \bibinfo{title}{{Agent-based spin model for financial markets on
  complex networks: Emergence of two-phase phenomena}}, \bibinfo{journal}{Phys.
  Rev. E} \bibinfo{volume}{78}~(\bibinfo{number}{3}) (\bibinfo{year}{2008})
  \bibinfo{pages}{1--6}, ISSN \bibinfo{issn}{1539-3755},
  \doi{\bibinfo{doi}{10.1103/PhysRevE.78.036115}}.

\bibitem[{Friedman and Rust(1993)}]{Friedman1993}
\bibinfo{author}{D.~Friedman}, \bibinfo{author}{J.~Rust}, \bibinfo{title}{The
  Double Auction Market}, \bibinfo{publisher}{The Advanced Book Program},
  \bibinfo{year}{1993}.

\bibitem[{Johnson(2010)}]{Johnson2010}
\bibinfo{author}{B.~Johnson}, \bibinfo{title}{Algorithmic Trading \& DMA},
  \bibinfo{publisher}{4Myeloma Press}, \bibinfo{year}{2010}.

\bibitem[{Chakraborti et~al.(2011)Chakraborti, Toke, Patriarca, and
  Abergel}]{Chakraborti2011}
\bibinfo{author}{A.~Chakraborti}, \bibinfo{author}{I.~M. Toke},
  \bibinfo{author}{M.~Patriarca}, \bibinfo{author}{F.~Abergel},
  \bibinfo{title}{{Econophysics Review: I. Empirical facts}},
  \bibinfo{journal}{Quant. Finance} \bibinfo{volume}{11}~(\bibinfo{number}{7})
  (\bibinfo{year}{2011}) \bibinfo{pages}{991--1012}.

\bibitem[{Mandelbrot(1963)}]{Mandelbrot1963}
\bibinfo{author}{B.~Mandelbrot}, \bibinfo{title}{{The variation of certain
  speculative prices}}, \bibinfo{journal}{The Journal of Business}
  \bibinfo{volume}{36}~(\bibinfo{number}{4}) (\bibinfo{year}{1963})
  \bibinfo{pages}{394--419}.

\bibitem[{Plerou et~al.(1999)Plerou, Gopikrishnan, {Nunes Amaral}, Meyer, and
  Stanley}]{Plerou1999}
\bibinfo{author}{V.~Plerou}, \bibinfo{author}{P.~Gopikrishnan},
  \bibinfo{author}{L.~A. {Nunes Amaral}}, \bibinfo{author}{M.~Meyer},
  \bibinfo{author}{H.~E. Stanley}, \bibinfo{title}{{Scaling of the distribution
  of price fluctuations of individual companies.}}, \bibinfo{journal}{Phys.
  Rev. E} \bibinfo{volume}{60}~(\bibinfo{number}{6 Pt A})
  (\bibinfo{year}{1999}) \bibinfo{pages}{6519--29}, ISSN
  \bibinfo{issn}{1063-651X}.

\bibitem[{Mantegna and Stanley(1999)}]{Mantegna1999}
\bibinfo{author}{R.~N. Mantegna}, \bibinfo{author}{H.~E. Stanley},
  \bibinfo{title}{{Introduction to Econophysics: Correlations and Complexity in
  Finance}}, \bibinfo{publisher}{Cambridge University Press},
  \bibinfo{year}{1999}.

\bibitem[{Plerou et~al.(2004)Plerou, Stanley, Gabaix, and
  Gopikrishnan}]{Plerou2004}
\bibinfo{author}{V.~Plerou}, \bibinfo{author}{H.~E. Stanley},
  \bibinfo{author}{X.~Gabaix}, \bibinfo{author}{P.~Gopikrishnan},
  \bibinfo{title}{{On the origin of power-law fluctuations in stock prices}},
  \bibinfo{journal}{Quant. Finance} \bibinfo{volume}{4}~(\bibinfo{number}{1})
  (\bibinfo{year}{2004}) \bibinfo{pages}{11--15}, ISSN
  \bibinfo{issn}{1469-7688}.

\bibitem[{Clark(1973)}]{Clark1973}
\bibinfo{author}{P.~K. Clark}, \bibinfo{title}{{A Subordinated Stochastic
  Process Model with Finite Variance for Speculative Prices}},
  \bibinfo{journal}{The Econometric Society}
  \bibinfo{volume}{41}~(\bibinfo{number}{1}) (\bibinfo{year}{1973})
  \bibinfo{pages}{135--155}.

\bibitem[{Arthur et~al.(1996)Arthur, Holland, LeBaron, Palmer, and
  Tayler}]{Arthur1996}
\bibinfo{author}{W.~B. Arthur}, \bibinfo{author}{J.~H. Holland},
  \bibinfo{author}{B.~LeBaron}, \bibinfo{author}{R.~Palmer},
  \bibinfo{author}{P.~Tayler}, \bibinfo{title}{{Asset Pricing Under Endogenous
  Expectations in an Artificial Stock Market}}, in: \bibinfo{editor}{W.~B.
  Arthur}, \bibinfo{editor}{S.~N. Durlauf}, \bibinfo{editor}{D.~{H. Lane}}
  (Eds.), \bibinfo{booktitle}{The Economy as an Evolving Complex System II},
  vol. \bibinfo{volume}{1001}, \bibinfo{publisher}{Addison-Wesley},
  \bibinfo{pages}{15--44}, \bibinfo{year}{1996}.

\bibitem[{Mandelbrot(1997)}]{Mandelbrot1997}
\bibinfo{author}{B.~Mandelbrot}, \bibinfo{title}{{Fractals and Scaling in
  Finance}}, \bibinfo{publisher}{Springer}, ISBN \bibinfo{isbn}{0387983635},
  \bibinfo{year}{1997}.

\bibitem[{Sornette(1998)}]{Sornette1998}
\bibinfo{author}{D.~Sornette}, \bibinfo{title}{{Multiplicative processes and
  power laws}}, \bibinfo{journal}{Phys. Rev. E}
  \bibinfo{volume}{57}~(\bibinfo{number}{4}) (\bibinfo{year}{1998})
  \bibinfo{pages}{4811--4813}.

\bibitem[{Lux and Marchesi(1999)}]{Lux1999}
\bibinfo{author}{T.~Lux}, \bibinfo{author}{M.~Marchesi},
  \bibinfo{title}{{Scaling and criticality in a stochastic multi-agent model of
  a financial market}}, \bibinfo{journal}{Nature}
  \bibinfo{volume}{397}~(\bibinfo{number}{February}).

\bibitem[{Challet et~al.(2000)Challet, Chessa, Marsili, and
  Zhang}]{Challet2000}
\bibinfo{author}{D.~Challet}, \bibinfo{author}{A.~Chessa},
  \bibinfo{author}{M.~Marsili}, \bibinfo{author}{Y.~C. Zhang},
  \bibinfo{title}{{From Minority Games to real markets}},
  \bibinfo{journal}{Quant. Finance} \bibinfo{volume}{1}~(\bibinfo{number}{1})
  (\bibinfo{year}{2000}) \bibinfo{pages}{9}.

\bibitem[{Farmer et~al.(2004)Farmer, Gillemot, Lillo, Mike, and
  Sen}]{Farmer2004}
\bibinfo{author}{J.~D. Farmer}, \bibinfo{author}{L.~Gillemot},
  \bibinfo{author}{F.~Lillo}, \bibinfo{author}{S.~Mike},
  \bibinfo{author}{A.~Sen}, \bibinfo{title}{{What really causes large price
  changes?}}, \bibinfo{journal}{Quant. Finance}
  \bibinfo{volume}{4}~(\bibinfo{number}{4}) (\bibinfo{year}{2004})
  \bibinfo{pages}{383--397}.

\bibitem[{Gabaix et~al.(2003)Gabaix, Gopikrishnan, and Plerou}]{Gabaix2003}
\bibinfo{author}{X.~Gabaix}, \bibinfo{author}{P.~Gopikrishnan},
  \bibinfo{author}{V.~Plerou}, \bibinfo{title}{{A theory of power-law
  distributions in financial market fluctuations}}, \bibinfo{journal}{Nature}
  \bibinfo{volume}{423}~(\bibinfo{number}{May}) (\bibinfo{year}{2003})
  \bibinfo{pages}{267--270}, \doi{\bibinfo{doi}{10.1038/nature01622.1.}}

\bibitem[{Farmer and Lillo(2004)}]{Farmer2004a}
\bibinfo{author}{J.~D. Farmer}, \bibinfo{author}{F.~Lillo}, \bibinfo{title}{{On
  the origin of power-law tails in price fluctuations}},
  \bibinfo{journal}{Quant. Finance} \bibinfo{volume}{4}~(\bibinfo{number}{1})
  (\bibinfo{year}{2004}) \bibinfo{pages}{C7--C11}, ISSN
  \bibinfo{issn}{1469-7688}, \doi{\bibinfo{doi}{10.1088/1469-7688/4/1/C01}}.

\bibitem[{Cohen et~al.(1983)Cohen, Maier, Schwartz, and Whitcomb}]{Cohen1983}
\bibinfo{author}{K.~J. Cohen}, \bibinfo{author}{S.~F. Maier},
  \bibinfo{author}{R.~A. Schwartz}, \bibinfo{author}{D.~K. Whitcomb},
  \bibinfo{title}{{A simulation model of stock exchange trading}},
  \bibinfo{journal}{Simulation} \bibinfo{volume}{41}~(\bibinfo{number}{5})
  (\bibinfo{year}{1983}) \bibinfo{pages}{181--191}, ISSN
  \bibinfo{issn}{0037-5497}, \doi{\bibinfo{doi}{10.1177/003754978304100502}}.

\bibitem[{Kim and Markowitz(1989)}]{Kim1989}
\bibinfo{author}{G.~Kim}, \bibinfo{author}{H.~Markowitz},
  \bibinfo{title}{{Investment Rules, Margin, And Market Volatility}},
  \bibinfo{journal}{J. Portfol. Manage.}
  \bibinfo{volume}{16}~(\bibinfo{number}{1}) (\bibinfo{year}{1989})
  \bibinfo{pages}{45--52}.

\bibitem[{Frankel and Froot(1988)}]{Frankel1988}
\bibinfo{author}{J.~A. Frankel}, \bibinfo{author}{K.~A. Froot},
  \bibinfo{title}{{Explaining the Demand for Dollars: International Rates of
  Return and the Expectations of Chartists and Fundamentalists}}, in:
  \bibinfo{editor}{R.~Chambers}, \bibinfo{editor}{P.~Paarlberg} (Eds.),
  \bibinfo{booktitle}{Agriculture, Macroeconomics, and the Exchange Rate},
  \bibinfo{publisher}{Westview Press}, \bibinfo{year}{1988}.

\bibitem[{Chiarella(1992)}]{Chiarella1992}
\bibinfo{author}{C.~Chiarella}, \bibinfo{title}{{The dynamics of speculative
  behaviour}}, \bibinfo{journal}{Annals of Operations Research}
  \bibinfo{volume}{37}~(\bibinfo{number}{1}) (\bibinfo{year}{1992})
  \bibinfo{pages}{101--123}, ISSN \bibinfo{issn}{0254-5330}.

\bibitem[{Beltratti and Margarita(1993)}]{Beltratti1993}
\bibinfo{author}{A.~Beltratti}, \bibinfo{author}{S.~Margarita},
  \bibinfo{title}{{Evolution of trading strategies among heterogeneous
  artificial economic agents}}, in: \bibinfo{editor}{J.-A. Meyer},
  \bibinfo{editor}{H.~L. Roitblat}, \bibinfo{editor}{S.~W. Wilson} (Eds.),
  \bibinfo{booktitle}{From animals to animats 2}, \bibinfo{pages}{494--501},
  \bibinfo{year}{1993}.

\bibitem[{Levy et~al.(1994)Levy, Levy, and Solomon}]{Levy1994}
\bibinfo{author}{M.~Levy}, \bibinfo{author}{H.~Levy},
  \bibinfo{author}{S.~Solomon}, \bibinfo{title}{{A microscopic market model of
  the stock market}}, \bibinfo{journal}{Econ. Letters} \bibinfo{volume}{45}
  (\bibinfo{year}{1994}) \bibinfo{pages}{103--111}.

\bibitem[{Lux(1997)}]{Lux1997}
\bibinfo{author}{T.~Lux}, \bibinfo{title}{{Time variation of second moments
  from a noise trader/infection model}}, \bibinfo{journal}{J. of Econ. Dyn.
  Control} \bibinfo{volume}{22}~(\bibinfo{number}{1}) (\bibinfo{year}{1997})
  \bibinfo{pages}{1--38}, ISSN \bibinfo{issn}{01651889},
  \doi{\bibinfo{doi}{10.1016/S0165-1889(97)00061-4}}.

\bibitem[{Mike and Farmer(2008)}]{Mike2008}
\bibinfo{author}{S.~Mike}, \bibinfo{author}{J.~D. Farmer}, \bibinfo{title}{{An
  empirical behavioral model of liquidity and volatility}},
  \bibinfo{journal}{J. of Econ. Dyn. Control}
  \bibinfo{volume}{32}~(\bibinfo{number}{1}) (\bibinfo{year}{2008})
  \bibinfo{pages}{200--234}, ISSN \bibinfo{issn}{01651889},
  \doi{\bibinfo{doi}{10.1016/j.jedc.2007.01.025}}.

\bibitem[{Gu and Zhou(2009)}]{Gu2009}
\bibinfo{author}{G.-F. Gu}, \bibinfo{author}{W.-X. Zhou},
  \bibinfo{title}{{Emergence of long memory in stock volatility from a modified
  Mike-Farmer model}}, \bibinfo{journal}{Europhys. Lett.}
  \bibinfo{volume}{86}~(\bibinfo{number}{4}) (\bibinfo{year}{2009})
  \bibinfo{pages}{48002}, ISSN \bibinfo{issn}{0295-5075},
  \doi{\bibinfo{doi}{10.1209/0295-5075/86/48002}}.

\bibitem[{LeBaron(2002)}]{LeBaron2002}
\bibinfo{author}{B.~LeBaron}, \bibinfo{title}{{Building the Santa Fe artificial
  stock market}}, \bibinfo{year}{2002}.

\bibitem[{Ehrentreich(2007)}]{Ehrentreich2007}
\bibinfo{author}{N.~Ehrentreich}, \bibinfo{title}{{Agent-based modeling: The
  Santa Fe Institute artificial stock market model revisited}},
  \bibinfo{publisher}{Springer}, ISBN \bibinfo{isbn}{9783540738787},
  \bibinfo{year}{2007}.

\bibitem[{Schmitt et~al.(2012)Schmitt, Sch\"{a}fer, M\"{u}nnix, and
  Guhr}]{Schmitt2012}
\bibinfo{author}{T.~A. Schmitt}, \bibinfo{author}{R.~Sch\"{a}fer},
  \bibinfo{author}{M.~C. M\"{u}nnix}, \bibinfo{author}{T.~Guhr},
  \bibinfo{title}{{Microscopic understanding of heavy-tailed return
  distributions in an agent-based model}}, \bibinfo{journal}{Europhys. Lett.}
  \bibinfo{volume}{100} (\bibinfo{year}{2012}) \bibinfo{pages}{38005}.

\bibitem[{T\"or\"ok et~al.(2013)T\"or\"ok, Iniguez, Yasseri, San~Miguel, Kaski,
  and Kertesz}]{Torok2013}
\bibinfo{author}{J.~T\"or\"ok}, \bibinfo{author}{G.~Iniguez},
  \bibinfo{author}{T.~Yasseri}, \bibinfo{author}{M.~San~Miguel},
  \bibinfo{author}{K.~Kaski}, \bibinfo{author}{J.~Kertesz},
  \bibinfo{title}{Opinions, Conflicts, and Consensus: Modeling Social Dynamics
  in a Collaborative Environment}, \bibinfo{journal}{Phys. Rev. Lett.}
  \bibinfo{volume}{110}~(\bibinfo{number}{088701}),
  \doi{\bibinfo{doi}{10.1103/PhysRevLett.110.088701}}.

\end{thebibliography}

\end{document}